# $C_3$-Symmetry-induced Antisymmetric Planar Hall effect and Magnetoresistance in Single-Crystalline Ferromagnets

W. J. Qin, K. Y. Wang, B. Yang, Y. Z. Tian, B. W. Zheng, B. Y. Huang, Y. B. Yang, W. Q. Zou, D. Wu,* P. Wang†

*National Laboratory of Solid State Microstructures, Jiangsu Provincial Key Laboratory for Nanotechnology, Jiangsu Physical Science Research Center, Institute of atom manufacturing, Collaborative Innovation Center of Advanced Microstructures and Department of Physics, Nanjing University, Nanjing 210093, People's Republic of China*

**ABSTRACT**. The planar Hall effect (PHE) is typically symmetric under magnetic field reversal, as required by the Onsager reciprocity relations. Recent advances have identified the antisymmetric PHE as an intriguing extension in magnetic systems. While new mechanisms have been proposed, the role of conventional anisotropic magnetoresistance (AMR) in this phenomenon remains unclear. Here, we report the experimental discovery of an antisymmetric PHE and magnetoresistance, with respect to both magnetic field and magnetization, in single-crystal $Co_{30}Pt_{70}$ (111) thin films with $C_3$ rotational symmetry and perpendicular magnetic anisotropy (PMA). We demonstrate that both antisymmetric effects arise naturally from the intrinsic fourth-rank AMR tensor inherent to $C_3$-symmetric planes, assisted by PMA. Our findings link conventional AMR to antisymmetric galvanomagnetic responses, offering new insights into symmetry-governed transport in crystalline ferromagnets.

The planar Hall effect (PHE), characterized by a transverse voltage induced by coplanar electric ($E$) and magnetic ($H$) fields, serves as a valuable probe in topological matter research [1-7] and finds practical applications in magnetic sensing [8-10]. In Weyl/Dirac semimetals, the PHE primarily arises from the chiral anomaly [1-5], whereas in magnetic systems it stems mainly from anisotropic magnetoresistance (AMR) [11-14]. Despite its name, the PHE is essentially an off-diagonal magnetoresistance (MR) encoded in the symmetric dissipative part ($\rho_{ij} = \rho_{ji}$) of the resistivity tensor $\rho_{ij}$ [15,16]. It differs fundamentally from the in-plane Hall effect (IPHE) [17-21], though both phenomena produce transverse voltages under coplanar $E$ and $H$ fields. The IPHE, as a genuine Hall effect, manifests in the antisymmetric non-dissipative part ($\rho_{ij} = -\rho_{ji}$) of $\rho_{ij}$. Constrained by Onsager reciprocity relations [22], the PHE is symmetric under magnetic field reversal, such that $\rho_{xy}(H) = \rho_{xy}(-H)$, while the IPHE is antisymmetric in $H$, satisfying $\rho_{xy}(H) = -\rho_{xy}(-H)$. This strictly holds for non-magnetic systems. Recent studies in magnetic systems with broken time-reversal symmetry have gone beyond this paradigm, discovering an unusual manifestation of the PHE termed the antisymmetric PHE [23]. This effect is antisymmetric with respect to $H$ despite its magnetoresistive origin and is emerging as a focus of interest in magnetic Weyl semimetals and magnetic heterostructures [24-27].

In magnetic systems with spontaneous magnetization $M$, Onsager's relations take the form $\rho_{ij}(H, M) = \rho_{ji}(-H, -M)$, which imposes a constraint only when $H$ and $M$ are reversed simultaneously. When $H$ and $M$ are treated as independent, an $H$-antisymmetric PHE becomes theoretically permitted upon $H$ reversal with fixed $M$, and additionally, PHE responses antisymmetric in both $H$ and $M$ are also allowed, with $\rho_{xy}(H, M) = -\rho_{xy}(-H, M) = -\rho_{xy}(H, -M)$. These theoretical possibilities have motivated recent experimental efforts to explore antisymmetric PHE in magnetic systems and they are attributed to new underlying mechanisms. For instance, in the magnetic Weyl semimetal $Co_3Sn_2S_2$ [24], the observed Hall response antisymmetric in both $H$ and $M$ is attributed to the interplay of the Berry curvature, the tilt of Weyl nodes and the chiral anomaly, while in the magnetic heterostructure CuPt/CoPt [26], similar effect is linked to the trigonal warping of the Fermi surface.

Although new mechanisms for the observed antisymmetric PHE in magnetic systems are being proposed, the contribution of conventional AMR to this phenomenon remains poorly understood. In isotropic polycrystalline ferromagnets [8,9,28,29], the well-established AMR-driven PHE expression is $\rho_{xy} \propto m_x m_y$, where $m_x$ and $m_y$ are the in-plane components of the normalized magnetization vector

*dwu@nju.edu.cn
†pengwang@nju.edu.cn

**m**. Since in-plane magnetic fields typically induce simultaneous sign reversal in $m_x$ and $m_y$, this configuration cannot produce the antisymmetric PHE phenomenon. However, in single-crystal ferromagnets, crystal symmetry governs the AMR tensor, leading to more complex PHE behaviors [30-38]. For instance, in systems with $C_4$ rotational symmetry, such as the (001) plane of cubic single-crystal magnetic films, terms like $m_x^2$ and $m_y^2$ in $\rho_{xy}$ have recently been experimentally confirmed in PHE measurements [39,40]. In lower-symmetry planes like the cubic (111) plane with $C_3$ symmetry, phenomenological theory indicates the emergence of new cross-terms in $\rho_{xy}$, such as $m_x m_z$ and $m_y m_z$, which couple in-plane and out-of-plane magnetization components (see Supplementary Material [41]). Under magnetic anisotropy, these new terms may generate antisymmetric PHE, a possibility that prior studies have largely overlooked.

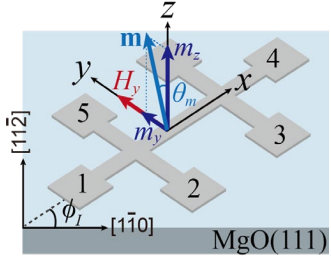

FIG. 1. Schematic of antisymmetric PHE measurement. A Hall bar device is patterned from a (111)-oriented magnetic film with strong PMA, epitaxially grown on MgO(111). An *xyz* coordinate system is established on the Hall bar, with the *x*-axis aligned along the device's longitudinal axis. $\phi_I$ denotes the angle between the longitudinal axis and the $[1\bar{1}0]$ crystallographic direction. $\theta_m$ denotes the angle between *z*-axis and **m**. Contacts are labeled by 1-5 to define transverse resistivities $\rho_{xy}$ and $\rho_{yx}$, enabling the distinction of PHE from the genuine Hall effect via coordinate-exchange symmetry.

Consider a single-crystal ferromagnetic film exhibiting the $C_3$-symmetry-induced $m_x m_z$ and $m_y m_z$ terms in $\rho_{xy}$, along with strong perpendicular magnetic anisotropy (PMA), as schematized shown in Fig. 1. Upon application of a relatively small in-plane field $H_y$, the magnetization **m** tilts towards the *y*-direction, resulting in $m_y \propto H_y$ and $m_z \approx \text{const}$. Due to PMA and small $H_y$. Consequently, the PHE measurement gives $\rho_{xy} \propto m_y m_z \propto H_y m_z$. This implies that $\rho_{xy}$ reverses sign upon reversal of either $H_y$ or $m_z$, manifesting a PHE response that is antisymmetric with respect to both $H_y$ and $m_z$—closely resembling previously reported phenomena [24,26]. Similarly, an in-plane magnetic field in the *x*-direction would induce analogous behavior through the $m_x m_z$ term. Despite the theoretical plausibility, experimental verification of this $C_3$-symmetry-induced antisymmetric PHE and MR mechanism remains unexplored.

In this work, we report the experimental discovery of an antisymmetric PHE in single-crystal ferromagnetic CoPt (111) thin films with both $C_3$ rotational symmetry and PMA. The observed PHE shows antisymmetric in both $H$ and $M$, manifesting threefold symmetry with varying $\phi_I$. A corresponding antisymmetric MR is also detected. Through phenomenological theory, we derived unconventional cross-terms $m_x m_z$ and $m_y m_z$ in both the transverse ($\rho_{xy}$) and longitudinal ($\rho_{xx}$) resistivities for the measurement in the cubic (111) plane. These terms are further experimentally verified via angle-dependent magnetoresistance (ADMR) measurements. Quantitative analysis of antisymmetric PHE, MR, and ADMR coefficients confirms that the antisymmetric PHE originates from the AMR cross-terms, assisted by PMA. Our results provide new insights into the origin of the antisymmetric PHE in single-crystal magnetic systems.

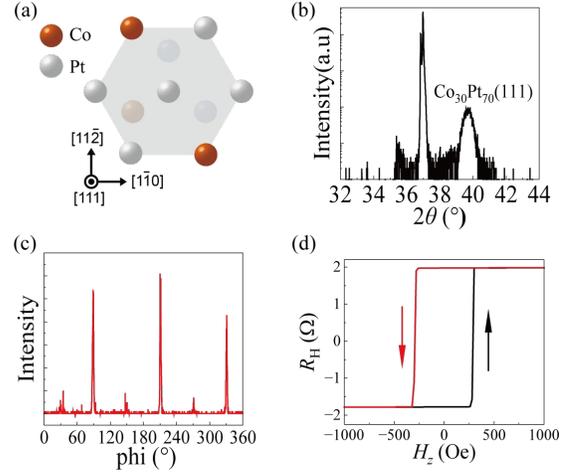

FIG. 2. (a) The plane view of the disordered cubic A1-phase (111)-oriented CoPt film. (b) $\theta$-$2\theta$ XRD pattern of the CoPt film epitaxially grown on the MgO(111). (c) Phi-scan XRD pattern with the CoPt(001) plane rotated along [111] axis, confirming the $C_3$ rotational symmetry. (d) Hall resistance $R_H$ measured with out-of-plane field $H_z$.

The epitaxial $Co_{30}Pt_{70}$ (CoPt) single-crystalline thin films were co-deposited on MgO (111) substrates at 300 °C via DC magnetron sputtering with Co and Pt deposition powers fixed at 12 W and 30 W respectively. Figure 2(a) presents the plane view of the (111)-oriented CoPt structure. Figure 2(b) shows the

θ-2θ X-ray diffraction (XRD) pattern of the epitaxial CoPt film on MgO (111), revealing a prominent CoPt (111) diffraction peak. A broad-range θ-2θ measurement indicates that the epitaxial CoPt adopts a disordered cubic A1-phase (see Supplementary Material [41]), where Co and Pt atoms are randomly distributed on lattice sites [42]. The (111) plane of this structure globally preserves $C_3$ rotational symmetry, as confirmed by the phi-scan measurements, that show three dominant CoPt(001) peaks in Fig. 2(c) [43,44]. Although three minor peaks suggest the presence of some twinned crystallites, the $C_3$ symmetry remains dominant.

The films were patterned into Hall bar devices via photolithography and ion beam etching, with orientations aligned along different azimuth angles (denotes by $\phi_I$) relative to the $[1\bar{1}0]$ crystallographic direction, as illustrated in Fig. 1. An xyz coordinate system is defined on the Hall bar, with the x-axis aligned along the longitudinal axis of the Hall bar. The contacts were labeled by 1-5 to define transverse resistivity $\rho_{xy}$ and $\rho_{yx}$. Specifically, we denote the two cases $(I_{14}, V_{25})$ and $(I_{25}, V_{14})$ as $\rho_{xy}$ and $\rho_{yx}$, respectively. This notation enables the distinction between the PHE and the genuine Hall effect though the symmetry under exchange of two coordinates. In particular, the PHE is symmetric under the subscript exchange $(\rho_{xy}^{PHE} = \rho_{yx}^{PHE})$, hence $\rho_{xy}^{PHE} = (\rho_{xy} + \rho_{yx})/2$, whereas the genuine Hall component is antisymmetric $(\rho_{xy}^{Hall} = -\rho_{yx}^{Hall})$, hence $\rho_{xy}^{Hall} = (\rho_{xy} - \rho_{yx})/2$. All the transport measurements were performed at 300 K. Figure 2(d) shows the Hall resistance $R_H$ measured with applying out-of-plane magnetic field $H_z$. The sharp switching of the loop with the coercivity field of ~ 250 Oe reveals the strong PMA of the CoPt films.

Figure 3 presents transverse resistivity measurements for a 5.5-nm CoPt (111) sample. The sample was first pre-magnetized along $+z$ ($-z$) direction (denoted by $m+$ and $m-$), after which $\rho_{xy}$ and $\rho_{yx}$ were recorded during the in-plane magnetic field $H_y$ swept up to 500 Oe. Figures 3(a) and 3(b) display the raw data of the measured $\rho_{xy}$ (solid squares) and $\rho_{yx}$ (hollow squares) as a function of $H_y$ at $\phi_I = 0°$ and 90°, respectively, for both $m+$ (red) and $m-$ (black) states. Although all curves exhibit quadratic behaviors, consistent with the anomalous Hall effect (AHE) in magnetic systems with strong PMA [45], the peaks/dips of the curves are shifted from $H_y = 0$. This deviation suggesting an additional contribution beyond pure AHE behavior.

To further analyze $\rho_{xy}$, we decompose $\rho_{xy}$ into contributions from the PHE ($\rho_{xy}^{PHE}$) and AHE ($\rho_{xy}^{AHE}$) using $\rho_{xy}^{PHE} = (\rho_{xy} + \rho_{yx})/2 - \rho_{bg}$ and $\rho_{xy}^{AHE} = (\rho_{xy} - \rho_{yx})/2$, respectively, based on their distinct symmetries of these two effects under exchanging coordinates [15,16]. The background signal $\rho_{bg} = [\rho_{xy}(H = 0) + \rho_{yx}(H = 0)]/2$ corresponds to the longitudinal magnetoresistance that leaks into the transverse channel and is therefore removed. As shown in Figs. 3(c) and 3(d), $\rho_{xy}^{PHE}$ is very different for $\phi_I = 0°$ and 90°, while $\rho_{xy}^{AHE}$ changes little (see Supplementary Material [41]). Crucially, at $\phi_I = 0°$, $\rho_{xy}^{PHE}$ exhibits a clear linear dependence on $H_y$, with its slope reversing sign upon switching the pre-magnetization direction. This behavior unequivocally identifies a linear PHE contribution that is antisymmetric with respect to both $H$ and $M$, which we denote as $\rho_{xy}^{PHE} \propto H_y m_z$, similar with the previous reports [24,26].

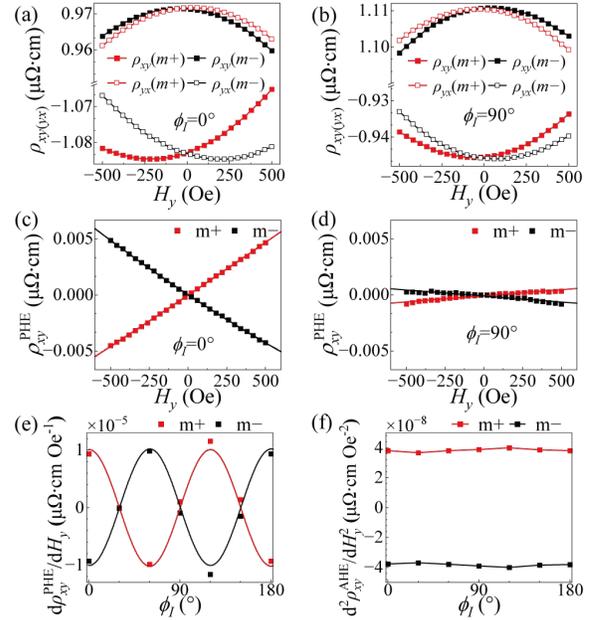

FIG. 3. Transverse resistivity measurements for $\phi_I = 0°$ (a) and 90° (b) respectively. The solid (hollow) squares correspond to $\rho_{xy}$ ($\rho_{yx}$) and the red (black) squares represent **m** pre-magnetized along $+z$ ($-z$). (c) $\rho_{xy}^{PHE}$ and (d) $\rho_{xy}^{PHE}$ extracted from (a) (b), respectively. The solid lines are the linear fitting through the origin. The dependence of (e) the slope of the $H$-antisymmetric PHE, $d\rho_{xy}^{PHE}/dH_y$, and (f) the AHE-derived second derivative, $d^2\rho_{xy}^{AHE}/dH_y^2$, on the Hall bar azimuth angle $\phi_I$. The solid line in (e) is the fitting with $\cos(3\phi_I)$.

Figure 3(e) plots the slope of the antisymmetric PHE, $d\rho_{xy}^{PHE}/dH_y$, as a function of the azimuthal angle $\phi_I$ ranging from 0° to 180°. The data exhibit a pronounced 120° periodicity, consistent with the $C_3$ rotational symmetry of the cubic (111) plane, wherein the crystal structure repeats identically every 120° upon rotation. Conversely, the AHE-derived second derivative $d^2\rho_{xy}^{AHE}/dH_y^2$ in Fig. 3(f) shows negligible $\phi_I$ dependence, as expected since the AHE (to first order in $m$) is isotropic under cubic symmetry. The good fitting for $d\rho_{xy}^{PHE}/dH_y$ with $\cos(3\phi_I)$ [solid line in Fig. 3(e)] allows one to have the relation of the PHE as $\rho_{xy}^{PHE} \propto H_y m_z \cos(3\phi_I)$.

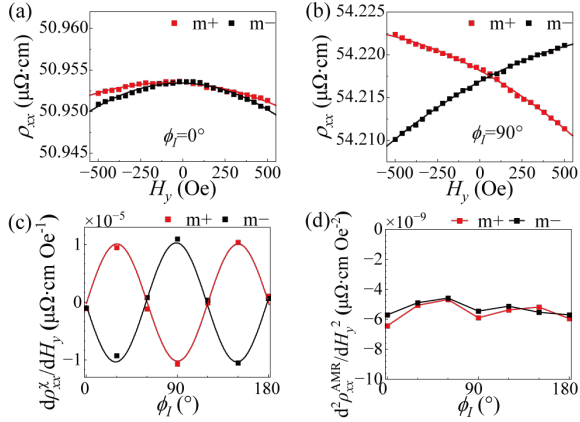

FIG. 4. Longitudinal resistivity measurements for $\phi_I = 0°$ (a) and 90° (b) respectively. The red (black) lines represent **m** pre-magnetized along $+z$ ($-z$). (c) The dependence of the slope of the antisymmetric linear MR, $d\rho_{xx}^\chi/dH_y$, on the Hall bar azimuth angle $\phi_I$. (d) $d^2\rho_{xx}^{AMR}/dH_y^2$ shows negligible dependence on $\phi_I$.

Next, we turn to the longitudinal resistivity measurements. Data for $\phi_I = 0°$ and 90° are presented in Figs. 4(a) and 4(b), respectively, with the sample pre-magnetized along $+z$ ($-z$) before $H_y$ sweeps. At $\phi_I = 0°$, the $\rho_{xx}$ versus $H_y$ curves exhibit the expected parabolic dependence for AMR, where $\rho_{xx}^{AMR} \propto m_y^2 \propto H_y^2$. The minor discrepancy between the $m+$ (red) and $m-$ (black) curves is likely due to differences in the slight misalignment of the magnetization from the $z$-axis between the two states, given that the AMR change in this range is less than $6 \times 10^{-5}$. In contrast, at $\phi_I = 90°$, the curves exhibit a significant difference. A dominant linear dependence of $\rho_{xx}$ on $H_y$ emerges, whose amplitude significantly exceeds the quadratic background. Notably, the slope of this linear dependence reverses sign when the pre-magnetization direction is reversed. This observation demonstrates a linear MR contribution that is antisymmetric with respect to both $H$ and $M$, which we denote as $\rho_{xx}^\chi \propto H_y m_z$, consistent with the previous report [23]. Moreover, similar phenomenon was observed in $Co_3Sn_2S_2$ [24,25]. To quantitatively analyze this behavior, we fit $\rho_{xx}$ using a combination of the quadratic AMR term $\rho_{xx}^{AMR}$ and the linear term $\rho_{xx}^\chi$, which represents the antisymmetric MR.

Figure 4(c) plots the slope of the antisymmetric MR, $d\rho_{xx}^\chi/dH_y$, as a function of the azimuthal angle $\phi_I$ ranging from 0° to 180°. The data exhibit a 120° periodicity similar to that of $d\rho_{xy}^{PHE}/dH_y$ shown in Fig. 3(e) but with a 90° phase shift. In contrast, the second derivative $d^2\rho_{xx}^{AMR}/dH_y^2$, arising from the conventional AMR term, shows negligible dependence on $\phi_I$ as depicted in Fig. 4(d). The expression for $\rho_{xx}^\chi$ can be narrowed down to $\rho_{xx}^\chi \propto H_y m_z \sin(3\phi_I)$. The nearly identical behavior compared to $\rho_{xy}^{PHE} \propto H_y m_z \cos(3\phi_I)$, differing only by a 90° phase shift, suggests that $\rho_{xx}^\chi$ and $\rho_{xy}^{PHE}$ share a common physical origin. This is further supported by the comparable magnitudes of $d\rho_{xx}^\chi/dH_y$ and $d\rho_{xy}^{PHE}/dH_y$, as shown in Fig. 4(c) and Fig. 3(e).

The observed antisymmetric PHE and MR originate from recently proposed mechanisms of Wely physics [2,24] can be immediately ruled out, since the ferromagnetic single-crystal films used in this study lack the specific electronic band structures required by such models. To interpret our experimental observations, we employ a phenomenological theory. Starting from a general form of the AMR tensor, we derive expressions for both $\rho_{xx}$ and $\rho_{xy}$ appropriate for the cubic (111) plane with $C_3$ symmetry.

For a cubic lattice (e.g., with point group m$\bar{3}$m), the fourth-order magnetoresistance tensor $a_{ijkl}$ ($\rho_{ij} = a_{ijkl} m_k m_l$) possesses only three independent elements due to the constraints of cubic symmetry. Defining the $XYZ$ coordinate system aligned with the orthogonal <001> principal axes, the electric field **E** in response to an applied current density **j** via AMR can be expressed as:

$$E_X = (a_1 m_X^2 + a_2 m_Y^2 + a_2 m_Z^2)j_X \\ + a_3 m_X m_Y j_Y + a_3 m_X m_Z j_Z \quad (1)$$

$$E_Y = (a_2 m_X^2 + a_1 m_Y^2 + a_2 m_Z^2)j_Y \\ + a_3 m_X m_Y j_X + a_3 m_Y m_Z j_Z \quad (2)$$

$$E_Z = (a_2 m_X^2 + a_2 m_Y^2 + a_1 m_Z^2)j_Z \\ + a_3 m_X m_Z j_X + a_3 m_Y m_Z j_Y \quad (3)$$

where $a_1, a_2, a_3$ are the three independent elements of the tensor $a_{ijkl}$ in the cubic lattice. By performing a coordinate rotation from the $XYZ$ coordinate system to the (111)-oriented $xyz$ coordinate system as shown in Fig. 1, we obtain the expressions for $\rho_{xx}$ and $\rho_{xy}$ in the (111) plane (see details in Supplementary Material [41]):

$$\rho_{xx} = b_1 m_x^2 + b_2 m_y^2 + b_3 m_z^2$$
$$+ b_4 m_z (m_x \cos(3\phi_I) - m_y \sin(3\phi_I)) \quad (4)$$

$$\rho_{xy} = b_0 m_z + b_5 m_x m_y$$
$$- b_4 m_z (m_x \sin(3\phi_I) + m_y \cos(3\phi_I)) \quad (5)$$

where $b_1$, $b_2$, $b_3$, $b_4$, and $b_5$ are parameters that depend on $a_1$, $a_2$, and $a_3$, and $\phi_I$ denotes the angle between the x-axis and the $[1\bar{1}0]$ crystallographic direction. To facilitate comparison with experimental results, we have introduced an additional $b_0 m_z$ term representing the AHE contribution to $\rho_{xy}$.

Notably, $\rho_{xx}$ and $\rho_{xy}$ in the (111) plane exhibit unconventional cross-terms involving $m_x m_z$ and $m_y m_z$. These unique cross-terms, which couple the in-plane and out-of-plane magnetization components, are permitted in the (111) plane due to the absence of 180° rotational symmetry. Such AMR cross-terms have been rarely investigated in previous experimental studies. Additionally, these cross-terms follow a $\cos(3\phi_I)$ or $\sin(3\phi_I)$ angular dependence, directly reflecting the $C_3$ symmetry of the (111) plane.

To experimentally verify these unconventional AMR cross-terms, we conducted ADMR measurements. Here, we take $\rho_{xx}$ given in Eq. (4) as an example for verification, measuring the angular dependence of $\rho_{xx}$ in the xy, yz, and xz planes under a 9 T magnetic field. According to Eq. (4), the expressions for $\rho_{xx}$ in these three planes are:

$$\rho_{xx}^{(xy)} = b_1 m_x^2 + b_2 m_y^2 \quad (6)$$
$$\rho_{xx}^{(yz)} = b_2 m_y^2 + b_3 m_z^2 - b_4 m_y m_z \sin(3\phi_I) \quad (7)$$
$$\rho_{xx}^{(xz)} = b_1 m_x^2 + b_3 m_z^2 + b_4 m_x m_z \cos(3\phi_I) \quad (8)$$

Notably, cross-terms are present in the yz and xz planes. At $\phi_I = 90°$, $\rho_{xx}$ simplifies to $b_2 m_y^2 + b_3 m_z^2 - b_4 m_y m_z$ in the yz plane and to $b_1 m_x^2 + b_3 m_z^2$ in the xz plane. In this configuration, the $b_4 m_y m_z$ cross-term appears only in the yz plane.

Figure 5(a) shows the ADMR measurement results of the three planes at $\phi_I = 90°$. In the xy-plane, $\rho_{xx}^{(xy)}$ exhibits a conventional AMR angular dependence, with maxima occurring when the magnetic field is aligned along the x-axis (0° or 180°) and minima along the y-axis (90° or 270°), consistent with Eq. (6). In contrast, in the yz-plane, the maxima of $\rho_{xx}^{(yz)}$ are clearly shifted leftward relative to the z-axis (0° or 180°), indicating an additional contribution beyond the conventional $\cos(2\theta_m)$ dependence arising from the $b_2 m_y^2 + b_3 m_z^2$ terms in Eq. (7). The $\rho_{xx}^{(yz)}$ data can be well fitted by the expression $\rho_0 + \Delta\rho_{xx}^{\cos} \cos(2\theta_m) + \Delta\rho_{xx}^{\sin} \sin(2\theta_m)$. The additional $\Delta\rho_{xx}^{\sin} \sin(2\theta_m)$ term originates from the cross-term $b_4 m_y m_z$ in Eq. (7), as the latter can be reformulated as $b_4 m^2 \sin(2\theta_m)/2$. Thus, the ADMR results for the yz-plane confirm the existence of the $b_4 m_y m_z$ cross-term. In the xz-plane, no shift of $\rho_{xx}^{(xz)}$ is observed, which is consistent with Eq. (8) at $\phi_I = 90°$. The deviation of the $\rho_{xx}^{(xz)}$ away from a simple $\cos(2\theta_m)$ function arises from contributions of higher-order $\cos(4\theta_m)$ AMR terms.

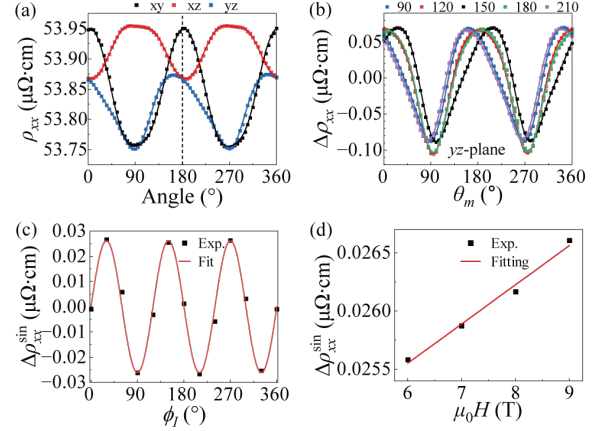

FIG. 5. (a) ADMR measurements at $\phi_I = 90°$ in the xy-, xz-, and yz-planes under a 9 T magnetic field. The solid lines are the fitted curves. (b) ADMR measurements in the yz-plane with $\phi_I$ from 90° to 210°. (c) Amplitude of the $\sin(2\theta_m)$ component versus azimuth angle $\phi_I$ extracted from the yz-plane ADMR data. (d) Linear fitting of the $\sin(2\theta_m)$ amplitude at $\phi_I = 90°$ under varying magnetic fields.

Having validated the existence of the $b_4 m_y m_z$ cross-term in Eq. (7), we proceed to characterize its dependence on the azimuthal angle $\phi_I$, specifically its $\sin(3\phi_I)$ dependence. Figure 5(b) presents yz-plane ADMR measurements with $\phi_I$ varied from 90° to 210°. The maxima of the ADMR curves shift systematically and exhibit a 120° periodicity, reflecting the underlying $C_3$ crystal symmetry. All curves in Fig. 5(b) were fitted using the expression: $\rho_0 + \Delta\rho_{xx}^{\cos} \cos(2\theta_m) + \Delta\rho_{xx}^{\sin} \sin(2\theta_m)$. The extracted $\Delta\rho_{xx}^{\sin}$ of the $\sin(2\theta_m)$ component—which originates from the $b_4 m_y m_z$ cross-term—is plotted as a function of $\phi_I$, as presented in Fig. 5(c). Notably, this dependence can be well fitted by a $\sin(3\phi_I)$ function. The confirmation of both the $b_4 m_y m_z$ and its $\sin(3\phi_I)$ dependence provide definitive evidence for the $b_4 m_y m_z \sin(3\phi_I)$ term in Eq. (7), thereby unambiguously confirming the existence of the unconventional AMR contributions.

We ascribe the observed antisymmetric PHE ($\rho_{xy}^{\text{PHE}}$) and MR ($\rho_{xx}^{\chi}$) signals in (111)-oriented CoPt films to the newly identified AMR cross-terms, assisted by the strong PMA. Within the measurement geometry employed, these signals specifically originate from the $b_4 m_y m_z \cos(3\phi_I)$ term in $\rho_{xy}$ and the $b_4 m_y m_z \sin(3\phi_I)$ term in $\rho_{xx}$, as indicated in Eqs. (4) and (5). As schematically illustrated in Fig. 1, a small magnetic field $H_y$ induces a linear magnetization response $m_y \approx \beta H_y$, yielding cross-terms of the form $\beta b_4 H_y m_z \cos(3\phi_I)$ and $\beta b_4 H_y m_z \sin(3\phi_I)$. These expressions match the experimental observations $\rho_{xy}^{\text{PHE}} \propto H_y m_z \cos(3\phi_I)$ and $\rho_{xx}^{\chi} \propto H_y m_z \sin(3\phi_I)$ precisely. Furthermore, all measured features of $\rho_{yx}$ and $\rho_{xx}$ shown in Figs. 2 and 3 are fully captured by Eqs. (4) and (5), which incorporate only the AMR and AHE contributions, without requiring any additional physical mechanisms.

To confirm the pure AMR origin of the antisymmetric PHE and MR, we compared the cross-term coefficient $b_4$ obtained from $\rho_{xy}^{\text{PHE}}$ and $\rho_{xx}^{\chi}$ measurements, which characterize the antisymmetric responses, with that derived from the ADMR measurements, which characterize the AMR terms. First, we estimated $b_4$ from the fitting results of $\rho_{xy}^{\text{PHE}}$ and $\rho_{xx}^{\chi}$ in Figs. 3(e) and 4(c). The magnetization tilt angle $\theta_m$ under $H_y$ = 500 Oe is determined via AHE fitting to be 7.93° (see Supplementary Material [41]). Using $b_4 m_y m_z = b_4 m^2 \sin(2\theta_m)/2$, we obtain $b_4$ = 0.040 μΩ·cm from $\rho_{xy}^{\text{PHE}}$ and $b_4$ = 0.042 μΩ·cm from $\rho_{xx}^{\chi}$. The discrepancy of less than 5% likely stems from experimental uncertainties.

Fitting the ADMR results in Fig. 5(c) yielded $b_4$ = 0.052 μΩ·cm, which is slightly larger than the values from the antisymmetric responses. This discrepancy can be explained by the different measurement conditions: ADMR is measured in the high-field regime, while the antisymmetric responses are measured at low fields, where effects such as incomplete magnetization alignment or a field-dependent magnetoresistance contribution [46] may play a role. To account for this, we conducted $yz$-plane ADMR measurements at various magnetic fields and performed linear extrapolations to isolate high-field contributions in Fig. 5(d). After removing the high-field contribution, the ADMR-derived $b_4$ is 0.046 μΩ·cm, which closely matches the $\rho_{xy}^{\text{PHE}}$ and $\rho_{xx}^{\chi}$ results, thereby confirming the pure AMR origin of the antisymmetric responses.

In summary, our experiments demonstrate that the observed antisymmetric PHE and MR in cubic CoPt fundamentally originates from the AMR mechanism. This phenomenon is fully captured by the fourth-rank magnetoresistance tensor without introducing ad hoc assumptions, where the presence of $C_3$ rotational symmetry plays a pivotal role. Moreover, considering that cubic crystals possess the highest point-group symmetry among crystalline materials, we anticipate that similar antisymmetric PHE and MR will manifest in a broad range of magnetic single-crystalline materials with strong magnetic anisotropy. Our work has clarified the most prevalent contribution of the antisymmetric PHE and MR in the single-crystal magnetic systems, and laid the foundation for further exploration of this intriguing phenomenon.

ACKNOWLEDGMENTS

This work was supported by National Key R&D Program of China (2022YFA1403602) and National Natural Science Foundation of China (52025012, 12334007, 12188101, 12374112), Natural Science Foundation of Jiangsu Province (BK20233001), Postdoctoral Fellowship Program of CPSF (GZB20230304) and China Postdoctoral Science Foundation (2023M741641).